\documentclass[aip,apl,groupedaddress,superscriptaddress,floatfix,showpacs,10pt,twocolumn]{revtex4-1}
\usepackage{graphicx}
\usepackage{color}

\begin{document}

\title{
Partially unzipped carbon nanotubes as magnetic field sensors
}

%

\author{S. Costamagna}
\affiliation{Universiteit Antwerpen, Department of Physics, Groenenborgerlaan 171, 2020 Antwerpen, Belgium.}
\affiliation{Facultad de Ciencias Exactas Ingenier{\'\i}a y Agrimensura, 
Universidad Nacional de Rosario and Instituto de F{\'\i}sica Rosario, Bv. 27 de Febrero 210 bis, 2000 Rosario,
 Argentina.}
%
\author{A. Schulz}
\author{L. Covaci}
%
\author{F. Peeters}
\affiliation{Universiteit Antwerpen, Department of Physics, Groenenborgerlaan 171, 2020 Antwerpen, Belgium.}

\date{\today}


\begin{abstract}
The conductance, $G(E)$, through graphene nanoribbons (GNR) connected to a partially unzipped carbon nanotube (CNT) 
is studied in the presence of an external magnetic field applied parallel to the long axis of the tube  
by means of non-equilibrium Green's function technique. We consider (z)igzag and (a)rmchair CNTs 
that are partially unzipped to form aGNR/zCNT/aGNR or zGNR/aCNT/zGNR junctions. 
We find that the inclusion of a longitudinal magnetic field affects the electronic states only in the CNT region, leading to the suppression 
of the conductance at low energies. 
Unlike previous studies, for the zGNR/aCNT/zGNR junction in zero field, we find a sharp dip in the conductance as 
the energy approaches the Dirac point and we attribute this non-trivial behavior to the peculiar 
band dispersion of the constituent subsystems. 
We demonstrate that both types of junctions can be used as magnetic field sensors.
\end{abstract}



\maketitle

\begin{figure}[t]
\includegraphics[height=1.1\columnwidth,angle=90]{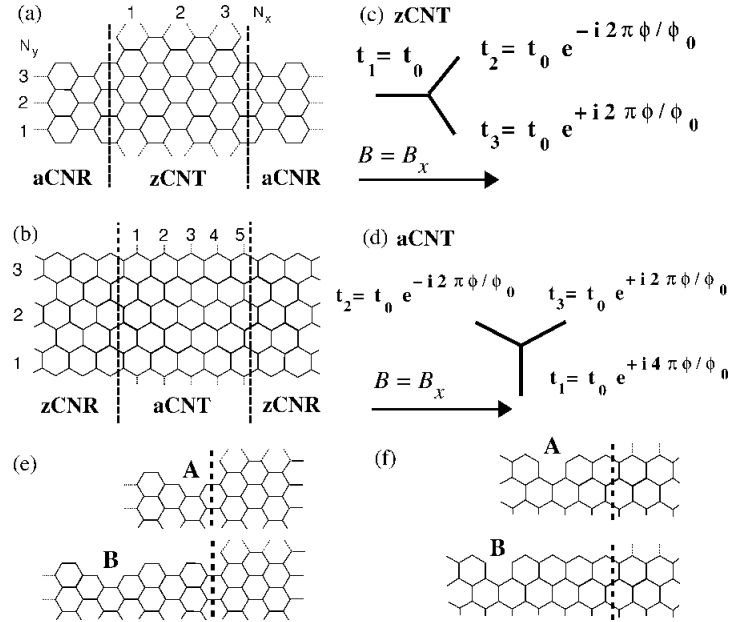}
\caption{
(a) aGNR/zCNT/aGNR and (b) zGNR/aCNT/zGNR junctions. 
The dashed thick vertical lines separate each part of the system considered for the conductance calculation. 
The Peierls factors (c-d) modifying the hopping amplitudes in the CNTs appearing due to the magnetic fields are 
$\phi=B r a \sqrt{3}/2$ for the zCNT and $\phi=B  r a/2$ for the zCNT. 
The size of the zCNT (aCNT) is defined by $N_x=3$ and $N_y=3$ ($N_x= 5$ and $N_y=3$) as indicated on the plot.
Edge vacancies considered in this work are displayed in (e) and (f). 
}
\label{fig1}
\end{figure}

%
%
Among the existing available techniques to produce GNRs\cite{ribb1,ribb2,ribb3,ribb4}, a very appealing one is the unzipping of CNTs. 
This can be done, for example, by using chemical attack\cite{chem1} or thought plasma beam etching\cite{plasma1}. 
Although recent molecular dynamics studies have shown that dangling bonds tend to re-bond 
on partially unzipped CNT, by fully hydrogenating the opened edges this self-healing behavior can be suppressed\cite{unzip1}.
Thus, it is feasible that multi-junctions composed of CNTs and GNRs 
can be fabricated in the near future.
From a theoretical point of view, the electronic transport properties of such junctions have been addressed
in a series of works. In Ref.~\onlinecite{brey1} the conductance of partially unzipped aCNT/zGNR junctions was numerically studied.  
The case of unzipped zCNT was addressed in Refs.~\onlinecite{Kly1,Kly2} where analytical expressions for the wave functions 
and transmission probabilities were obtained by connecting different parts of the system under proper boundary conditions. 
The main result is that the metallic aGNR/zCNT/aGNR junctions exhibit perfect transmission for incident low-energy electrons. 
Different from previous works, here we are interested to study the conductance of partially unzipped CNT in the presence of an 
external magnetic field parallel to the tube axis. 
We will analyze the aGNR/zCNT/zGNR and zGNR/aCNT/zGNR junctions schematically displayed in Fig.~\ref{fig1}.
We will show that variations in the external magnetic field can dramatically change the conductance of the system 
and therefore we propose to use these systems as magnetic field sensors. 
Such sensors have the additional advantage that much better contacts can be
realized on the ribbon sections as compared to plain nanotubes.
%

%
%
In order to describe the electron hopping between the $\pi$ orbitals pertaining to 
neighboring carbon atoms we adopt a nearest-neighbor tight binding Hamiltonian. 
The unzipped systems studied here (a) zCNT and (b) aCNT, are displayed in Fig.~\ref{fig1}.
Note that, for simplicity we have adopted a different enumeration $N_x$ and $N_y$ to define each geometry 
and that periodic boundary condition are used at the two edges to define the tubes.

\begin{figure}[t]
\includegraphics[width=0.98\columnwidth]{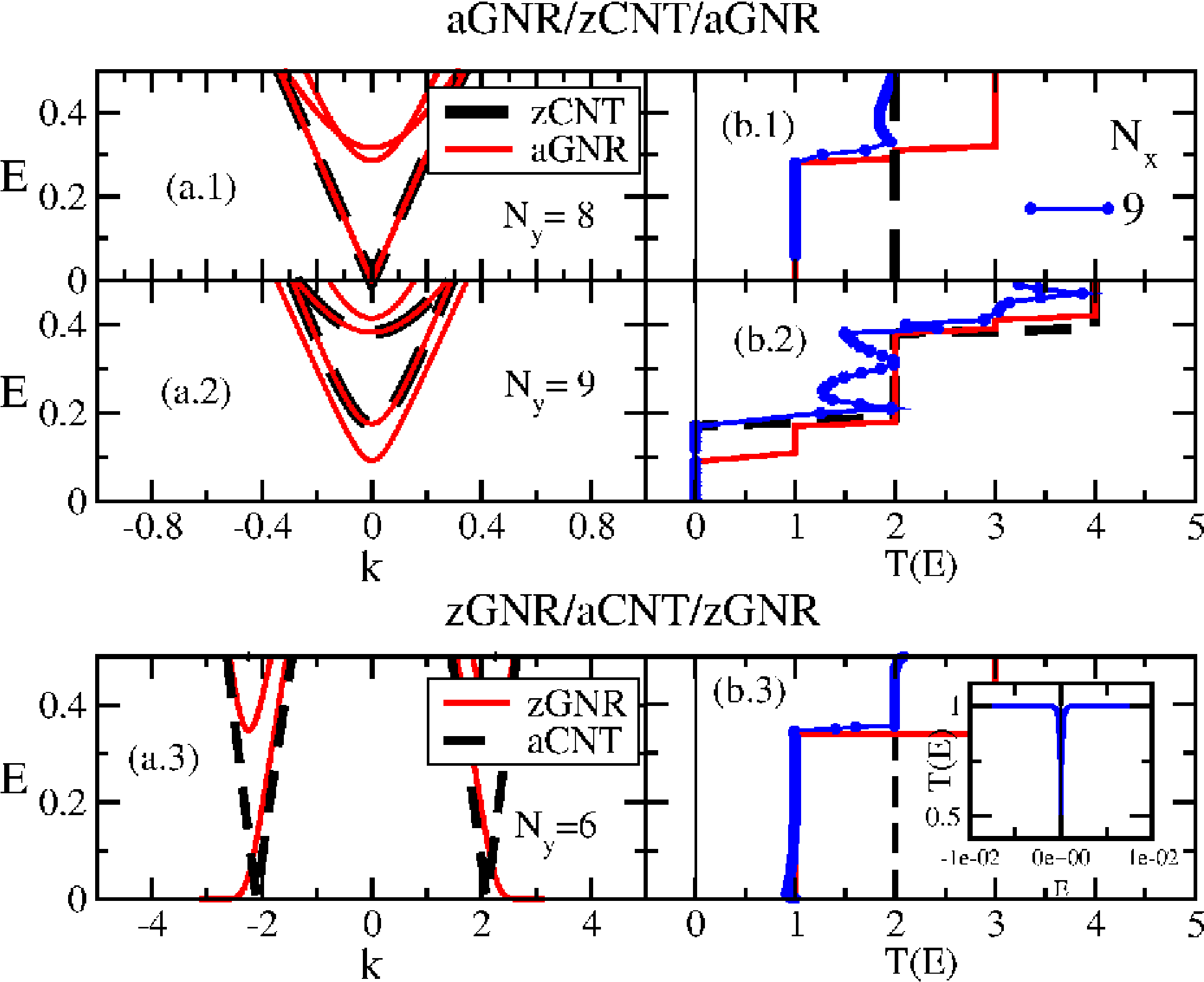}
\caption{(Color online)(a) Bandstructure of pure CNTs (dashed lines) and GNRs (solid lines) as indicated.
$T(E)$ for each corresponding junction is shown in (b) for $N_x=9$.
$T(E)$ for the pure CNTs (dashed lines) and GNR (solid lines) were included for a better comprehension.
The inset shows the vanishing of $T(E)$ at very low energies in the case of zGNR/aCNT/zGNR junctions (see text).
}
\label{fig2}
\end{figure}
When a CNT is partially unzipped we model it as a CNT that is connected to ribbons on both of its sides. 
The aGNR/zCNT/aGNR system can be thought of as a zCNT from which one line of C-atom dimers is removed in 
the left and right section, forming the aGNR-leads.  
In the zGNR/aCNT/zGNR system, no atoms are removed, but the bonds between two zigzag edges are removed.
In both cases the model Hamiltonian describing the whole system is therefore given by
\begin{equation}
H= H_{z(a)CNT}  + H_{a(z)GNR_L} 
+ H_{a(z)GNR_R} 
\label{hamil}
\end{equation}
 where $H_{a(z)GNR_{L(R)}}$ corresponds to the semi-infinite left (right) lead, and  
$H_{z(a)CNT}$ describes the nanotube. In addition to these terms we have $h_{LC}$ and $h_{LR}$ which are 
the hopping amplitudes connecting the leads to the nanotube.
The hopping parameter between carbon atoms, $t_0=2.7$ eV, is taken as unit of energy.
The conductance calculation was done by using the non equilibrium Green's function technique\cite{dattabook,xu,seba}
where $G(E)$ is obtained from the transmission function $T(E)$ as $G(E)=(2 e^2 /h) T(E)$.
We remark that the self-energies for the semi-infinite leads have been computed self-consistently and when used to calculate the $G(E)$ of pure CNTs and GNRs they give the expected value as shown in Fig.~\ref{fig2}.
%
%
%
Even without external magnetic field, very different low energy electronic properties can be present in these junctions. 
In the case of zCNTs/aGNRs, it is known that the metallicity of undoped samples is 
determined by the width of the tubes and ribbons.\cite{Kly1}
In Fig. \ref{fig2}, panels (a.1) and (a.2) show the low energy band-structure for pure infinitely long aGNRs and zCNTs of different widths. 
In both, the metallic and the semiconducting regimes, only some of the bands of the aGNR coincide with the 
bands of the zCNT due to the different transverse boundary condictions (open vs. periodic), see e.g. Ref. \onlinecite{Kly1}.
In addition, while each band of the aGNR is non-degenerate, in the zCNT the bands are doubly degenerate\cite{Kly1} which implies
that for low energies metallic zCNT possess a higher conductance than aGNR.
%
%
For the partially unzipped aGNR/zCNT/aGNR metallic junctions, i.e. $N_y=8$, it is expected that in the absence of the magnetic field 
the system should show no backscattering and a perfect conductance at low energies\cite{Kly1}.
Accordingly we see in Fig.~\ref{fig2} (b.1) that the transmission remains unity at low energies. 
For higher energies $G(E)$ shows the expected jumps each time that a new channel is available for conduction.

\begin{figure}[t]
\includegraphics[width=0.92\columnwidth]{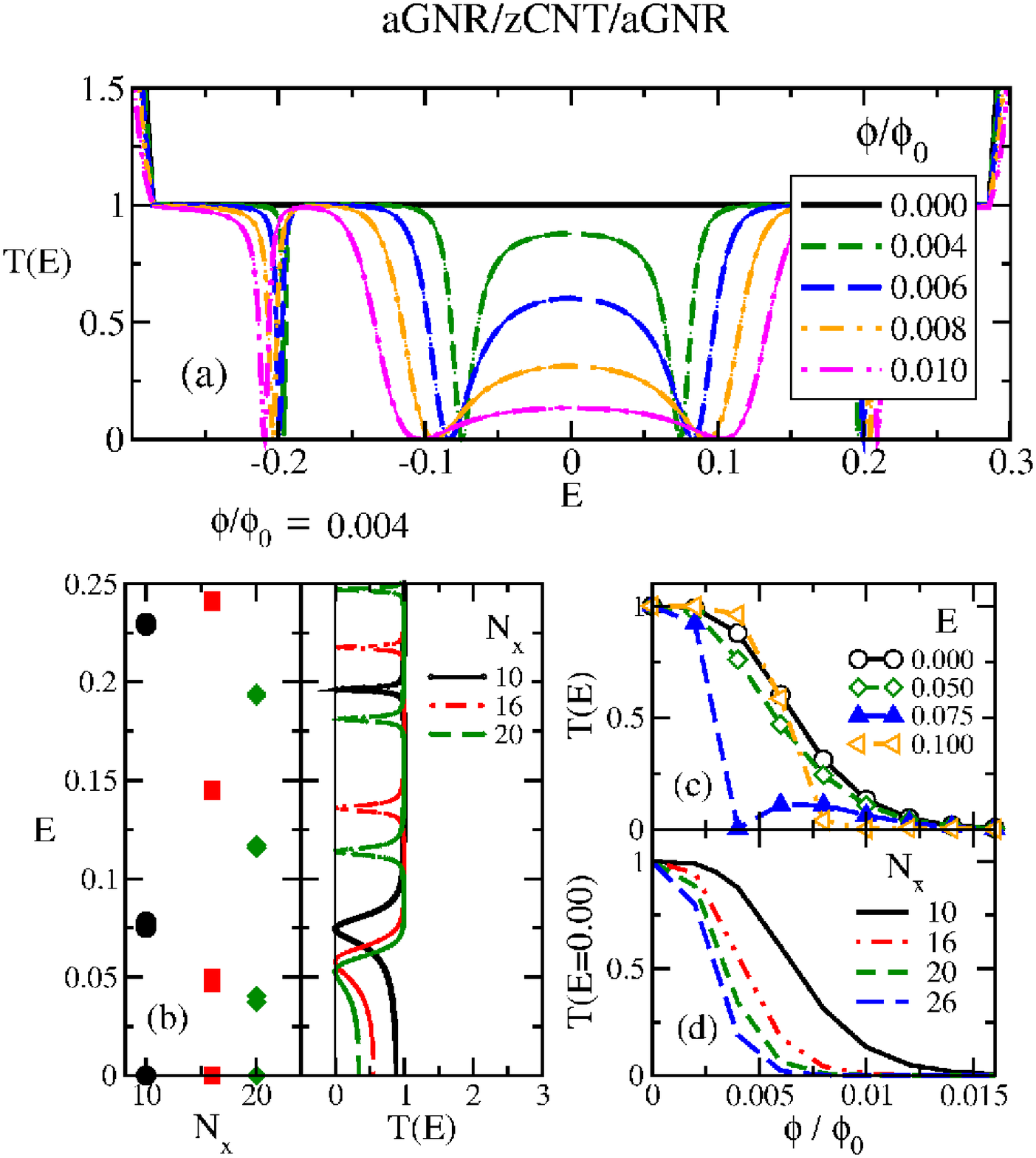}
\caption{(Color online) 
(a) $T(E)$ in the unzipped zCNT ($N_y=8$, $N_x=10$) for different magnetic fields.
(b) Low energy spectrum of an isolated zCNT and $T(E)$ for different tube lengths.
$\phi/\phi_0 =0.004$ and $N_y=8$ in both cases.
(c) Variation of $T(E)$ for several energies near zero (extracted from (a)) and
(d) decay of $T(E=0.00)$, for several tube length as indicated, against the magnetic field.
}
\label{fig3}
\end{figure}
%
%
On the other hand, in the case of the zGNR/aCNT/zGNR junction (Fig.~\ref{fig2} (a.3) and (b.3))
we observe a quite different low energy behavior. While zGNRs are always metallic independent of their width, 
due to the localized edge states at zero energy, $T(E)$ vanishes in the combined system when the energy approaches zero,
which is made more clear in the inset of Fig.~\ref{fig2}.
This sharp drop in the conductance at zero energy can be understood from the mismatch in the dispersion 
of the zGNR and aCNT subsystems, (cf. Fig.~\ref{fig2} (a.3)), as the energy approaches zero. 
Due to the resulting difference in momentum of the corresponding eigenstates, electron wave-functions incident from 
the ribbon become evanescent in the aCNT and therefore the conduction is suppressed. 
Although not shown here, we have found that this suppression is sensitive to the length of the aCNT subsystem, 
as the electron tunnels through it. 
%
%
%
%
%

\begin{figure}[th]
\includegraphics[width=0.9\columnwidth]{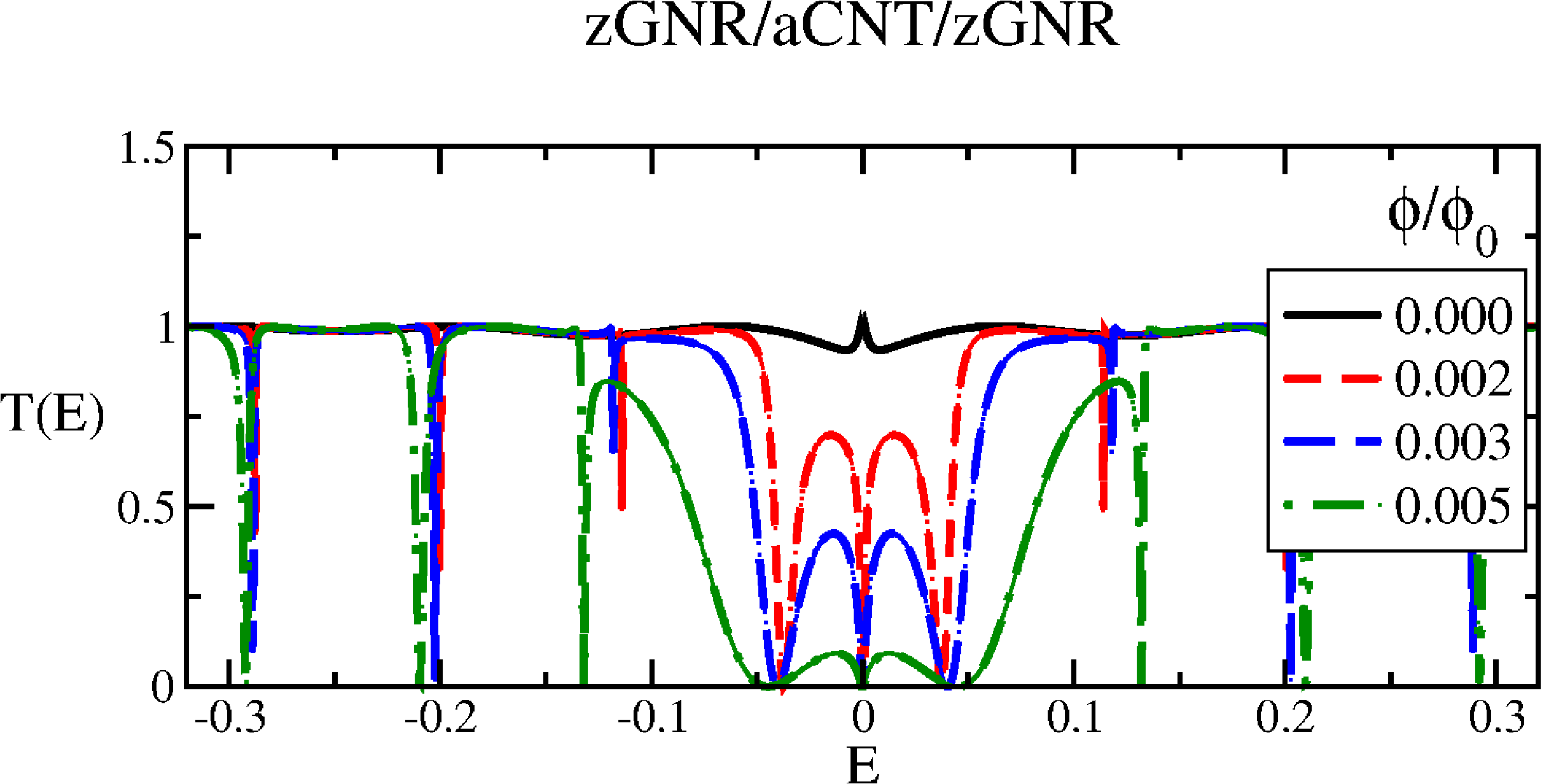}
\caption{(Color online)  Variation of $T(E)$ in the unzipped aCNT ($N_y=6$, $N_x=24$) 
for different magnetic fields as indicated on the plot. 
}
\label{fig4}
\end{figure}


When an external magnetic field is acting on the CNTs it can be included into the tight binding Hamiltonian 
using the so-called Peierls phase from which $t_{ij} = - t_0 e^{- i e/\hbar \int_i^j \vec{A}\cdot\vec{dl}}$
where $\vec{B}=\vec{\nabla} \times \vec{A}$ and the integration is done along the hopping path.
For a magnetic field parallel to the tube axis, this is $B=B_x$, there will 
be only three different hopping phases. These terms were displayed in
Fig.~\ref{fig1}(c) and Fig.~\ref{fig1}(d). $\phi_0=h/e$ is the quantum flux,
$r$ is the radius of the tube, $B$ is the magnitude of the magnetic field and 
$a=1.42 \AA$ is the carbon atoms distance. For the GNR the axial magnetic field has no effect since the flux through the ribbon is zero.

\begin{figure}[t]
\includegraphics[height=0.95\columnwidth,angle=90]{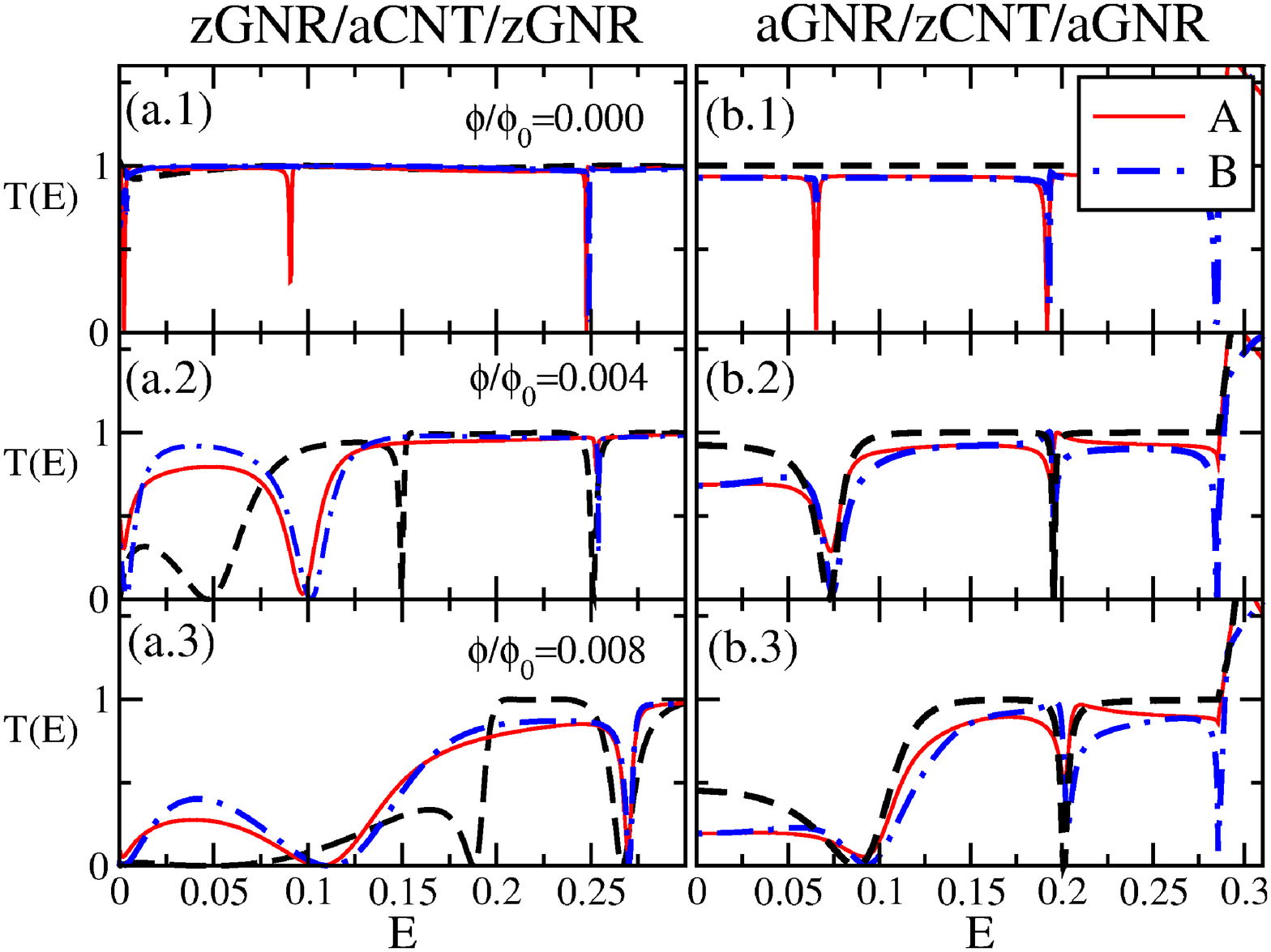}
\caption{(Color online) Effect of edge vacancies on $T(E)$ 
in the (a) zGNR/aCNT/zGNR ($N_x=10$, $N_y=6$) and (b) 
aGNR/ZCNT/aGNR ($N_x=10$, $N_y=8$) junctions 
for different magnetic fields as indicated. 
The location of the vacancies named A and B is displayed in Figs.~\ref{fig1}(e,f).
$T(E)$ for the case where there are no vacancies is indicated by the black-dashed line. 
}
\label{fig5}
\end{figure}

The effect of the magnetic field on the aGNR/zCNT/aGNR junction is shown in Fig.~\ref{fig3}. 
At low energies the transmittance is suppressed 
due to the magnetic field induced gap in the zCNT (this comes about because the magnetic field 
will shift the transverse momentum thus changing a metallic nanotube into a semiconducting one). 
The suppression is due to a decrease of the overlap of evanescent modes from the aGNR and  
depends on the tube length as can be seen in Fig.~\ref{fig3}(b).
Thus, when the transmittance vs. magnetic flux is plotted for different energies and tube lengths, as shown in Figs.~\ref{fig3}(c-d),  we notice that this junction could be used as a magnetic sensor since the resistance of the junction is very sensitive to magnetic field changes. 
The longer the tube length, more sensitive the sensor will be
since smaller magnetic fields are needed to suppress the overlap of the 
wave functions coming from the leads.
When the Fermi level is shifted from the Dirac point for certain energies, which depends on the tube length,  
the transmittance is even more sensitive to changes in the magnetic field thus improving the efficiency of the sensor.
For the zGNR/aCNT/zGNR junction, the dependence of the transmittance is presented in Fig.~\ref{fig4}. 
Except to the peculiar zero energy suppresion, the behavior is similar to the aGNR/ZCNT/aGNR junction 
thus making the operation of the proposed magnetic sensor independent of the orientation of the partially unzipped CNTs\cite{n11}.
%
The effect of defects as Carbon atoms vacancies on the edges of the ribbons, 
where they are more likely to appear in the unzipping process, is shown in Fig.~\ref{fig5}.
While edge defects modify the conductance, their effects can be clearly distinguished 
from the effects of the magnetic field and thereby the performance of these 
magnetic fields sensors holds.

Magnetic field induced changes of the electronic properties in the zCNT/aGNR regions will modify the magneto-resistance of the junctions which hence can be used as magnetic field sensors.
While more realistic system sizes are out of the scope of the numerical technique employed, 
given a tube radius of $r=500 nm$, $\phi/\phi_0$ equals to $0.001$ which implies a 
magnetic field  $B \approx 0.67-1.16$ T depending on the junction.
Therefore, the simplicity of the sensor, due to the fact that the GNRs are perfect contacts to the CNTs while the magnetic field does not affect the contacts, makes it interesting for possible applications.
Edge defects in the ribbons do not modify the qualitative behavior of the junctions
and their effect on the conduction can be neglected when the defects 
are far from the junction (e.g. farther than 1nm). 
Also, due to the extraordinary stiffness of CNTs, in the relaxed configuration 
the nanotube region is expected to be straight and therefore curvature effects can be safely neglected.
 
{\textit{Acknowledgments}}
LC acknowledges support from the Flemish Science Foundation (FWO-Vl) and SC from the Belgian Science Foundation (BELSPO).
This work is supported by the ESF-EuroGRAPHENE project CONGRAN.
\label{agradecimientos}


\begin{thebibliography}{99}
\bibitem{n11}Although not shown here, we have found the same behavior for metallic junctions with different widths.
\end{thebibliography}

\end{document}